\documentclass[prc,twocolumn,superscriptaddress]{revtex4}
\usepackage{graphicx}

\begin{document}

\title{Parton rescattering and screening in Au+Au collisions at RHIC}

\author{S. A. Bass}
\affiliation{Department of Physics, Duke University,
             Durham, NC 27708-0305, USA}
\affiliation{RIKEN BNL Research Center, Brookhaven National Laboratory, 
	Upton, NY 11973, USA}

\author{B. M\"uller}
\affiliation{Department of Physics, Duke University,
             Durham, NC 27708-0305, USA}

\author{D. K. Srivastava}
\altaffiliation{on leave from: Variable Energy Cyclotron Centre, 
	1/AF Bidhan Nagar, Kolkata 700 064, India}
\affiliation{Department of Physics, Duke University,
             Durham, NC 27708-0305, USA}

\begin{abstract}
We study the microscopic dynamics of quarks and gluons in relativistic
heavy ion collisions in the framework of the Parton Cascade Model.
We use lowest order perturbative QCD cross sections with fixed lower
momentum cutoff $p_T^{\rm min}$.
We calculate the time-evolution of the Debye-screening mass $\mu_D$
for Au+Au collisions at $\sqrt{s}=200$~GeV per nucleon pair. 
The screening mass
is used to determine a lower limit for the allowed
range of $p_T^{\rm min}$. 
We also determine the energy density reached through hard and 
semi-hard processes at RHIC, obtain a lower bound for the rapidity
density of charged hadrons produced by semihard interactions, and
analyze the extent of perturbative rescattering among partons.
\end{abstract}

\preprint{DUKE-TH-02-222}

\maketitle

The first experimental results from the Relativistic Heavy Ion
Collider (RHIC) have generated a vast amount of interesting 
data \cite{qm,reviews}.
A variety of theoretical models has been invoked to
describe the observed phenomena, e.g., fluid dynamical models,
perturbative QCD scattering models, as well as models based on 
parton saturation and statistical approaches.
Although these models, which all contain adjustable parameters,
have been fairly successful within their respective regimes of
anticipated applicability, they all have certain limitations. For
example, fluid dynamics cannot describe transport phenomena 
occurring prior to local equilibration of the produced matter, 
and it must fail above a certain, though unknown, value of $p_T$.
Perturbative parton scattering models cannot describe the
physics of equilibration and the formation of collective flow.
Initial state parton saturation models do not include the final 
state interaction among partons, which leads to equilibration.

The parton cascade model (PCM) was proposed, about a decade ago, with
the aim to provide a unified framework for the
description of phenomena involving high and intermediate values
of $p_T$ \cite{pcm_orig}. 
It is based on the premise that the microscopic
dynamics during the early stages of a high-energy nuclear collision 
can be described as a cascade of two-body interactions among
perturbative quarks and gluons, with certain modifications caused
by the presence of a hot and dense medium. Because of its
perturbative nature, the PCM cannot describe the hadronization 
stage, unless additional phenomenological models are introduced.

Since the original formulation and implementation of the PCM 
by Geiger (VNI) \cite{vni}, significant new insights into the dynamics of
dense systems of partons have been gained. First and foremost
among these is the applicability of semiclassical methods for 
the description of the partonic structure in the initial state and 
the earliest phase of its evolution \cite{mlv}. 
Another relevant
development is the recognition of the importance of radiative
processes for the attainment of local thermal equilibrium in
the rapidly expanding matter. Radiative processes were already
included in the original PCM in the leading-logarithmic 
approximation (LLA). The concept of initial-state saturation
of the parton distribution, however, provides a novel feature
which can be utilized to address the infrared divergences of
the perturbative PCM.

In this letter we focus on the early, pre-thermal reaction phase
of a heavy-ion reaction and address the following questions:
\begin{itemize}
\item What limits on the cut-off $p_T^{min}$, required by the 
	infra-red divergence of the pQCD cross section, are imposed 
	by the internal consistency
	of the model and by existing data on particle distributions?
\item What energy-density and multiplicity is reached through hard and 
	semi-hard processes (alone) at RHIC within these limits?
\item Do partons undergo multiple perturbative binary collisions? Are these
	sufficient to produce (measurable) collective effects?
\end{itemize}

We will here restrict our discussion of the results from the
Parton Cascade Model 
implementation (VNI/BMS, an improved and corrected version of the
VNI implementation \cite{vni}) to binary processes in leading 
order pQCD -- a more detailed calculation involving NLO corrections
such as radiative processes will be presented in a forthcoming 
publication \cite{bms_big1}.

The fundamental assumption underlying the PCM is that the state 
of the dense partonic system can be characterized by a set of 
one-body distribution functions $F_i(x^\mu,p^\alpha)$. Here $i$
denotes the flavor index ($i = g,u,\bar{u},d,\bar{d},\ldots$)
and $x^\mu, p^\alpha$ are coordinates in the eight-dimensional
phase space. The partons are assumed to be on their mass shell,
except before the first scattering. In the numerical implementation, the
continuous parton distribution functions (here we choose the GRV-HO 
parametrization \cite{grv}), are represented by (test) particles
\begin{equation}
F_i(x,\vec p) = \sum_{i=1}^N \int {\rm d} \tau \int {\rm d} 
p^0 \, \epsilon_i\,
\delta(x^\mu - \xi_i^\mu(\tau))\,
\delta(p^\alpha - q_i^\alpha(\tau))\; ,
\label{eq01}
\end{equation}
where $\xi_i^\mu(\tau)$ and $q_i^\alpha(\tau)$, respectively, 
denote the space-time position and four-momentum of particle $i$.
$\tau$ is a variable (proper time) parameterizing the world-line
of a particle. The factor $\epsilon_i = 0,1$ allows for the 
creation and annihilation of partons.

Partons generally propagate on-shell and on straight-line paths 
between scattering events.
Before their first collision, partons may have a space-like
four-momentum, especially if they are assigned an ``intrinsic''
transverse momentum.  

The time-evolution of the parton distribution is governed by a 
relativistic Boltzmann equation:
\begin{equation}
p^\mu \frac{\partial}{\partial x^\mu} F_i(x,\vec p) = {\cal C}_i[F]
\label{eq03}
\end{equation}
where the collision term ${\cal C}_i$ is a nonlinear functional 
of the phase-space distribution function. Although the collision
term, in principle, includes factors encoding the Bose-Einstein 
or Fermi-Dirac statistics of the partons, we neglect those effects
here.

The collision integrals have the form:
\begin{equation}
\label{ceq1}
{\cal C}_i[F] = \frac{(2 \pi)^4}{2 S_i E_i} \cdot
\int  \prod\limits_j {\rm d}\Gamma_j \, | {\cal M} |^2 
       \, \delta^4(P_{\rm in} - P_{\rm out}) \, 
          D(F_k(x, \vec p)) 
\end{equation}
with
\begin{equation}
D(F_k(x,\vec p)) \,=\, 
\prod\limits_{\rm out} F_k(x,\vec p) \, - \,
\prod\limits_{\rm in} F_k(x,\vec p) \quad
\end{equation}
and
\begin{equation}
\prod\limits_j {\rm d}\Gamma_j = \prod\limits_{{j \ne i} \atop {\rm  in,out}} 
        \frac{{\rm d}^3 p_j}{(2\pi)^3\,(2p^0_j)} 
\quad.   
\end{equation}
$S_i$ is a statistical factor defined as
$
S_i \,=\, \prod\limits_{j \ne i} K_a^{\rm in}!\, K_a^{\rm out}!
$
with $K_a^{\rm in,out}$ identical partons of species $a$ in the initial
or final state of the process, excluding the $i$th parton.

The matrix elements $| {\cal M} |^2$ account for the following 
processes:
\begin{equation}
\label{processes}
\begin{array}{lll}
g g \to g g \quad&\quad g g \to q \bar q \quad&\quad	q g \to q g \\
q q' \to q q' \quad&\quad q q \to q q \quad&\quad q \bar q \to q' \bar q' \\
q \bar q \to q \bar q \quad&\quad q \bar q \to g g \quad& \\
q g \to	q \gamma \quad&\quad q \bar q \to \gamma \gamma \quad & \quad q \bar q \to g \gamma
\end{array}
\end{equation}
with $q$ and $q'$ denoting different quark flavors.
The amplitudes for these processes have been calculated in refs. 
\cite{Cutler.78,Combridge.77} for massless quarks. The 
corresponding scattering cross sections are expressed in terms
of spin- and colour-averaged amplitudes $|{\cal M}|^2$:
\begin{equation}
\label{dsigmadt}
\left( \frac{{\rm d}\hat \sigma}
     {{\rm d} Q^2}\right)_{ab\to cd} \,=\, \frac{1}{16 \pi \hat s^2}
        \,\langle |{\cal M}|^2 \rangle
\end{equation}
For the transport calculation we also need the total cross section 
as a function of $\hat s$ which can be obtained from (\ref{dsigmadt}):
\begin{equation}
\label{sigmatot}
\hat \sigma_{ab}(\hat s) \,=\, 
\sum\limits_{c,d} \, \int\limits_{(p_T^{\rm min})^2}^{\hat s}
        \left( \frac{{\rm d}\hat \sigma }{{\rm d} Q^2}
        \right)_{ab\to cd} {\rm d} Q^2 \quad .
\end{equation}
The low momentum-transfer cut-off $p_T^{min}$ is needed
to regularize the IR-divergence of the  pQCD parton-parton cross section.
A novel feature of our treatment involves the introduction of a factor
$f(x,Q^2)/f(x,Q_0^2)$ for every primary parton involved in a binary
collision in expression~\ref{sigmatot} to account for the difference
between the initialization scale $Q_0$ and the scattering scale $Q$.
A more detailed description of our 
implementation is in preparation \cite{bms_big1}.

One of the most crucial parameters of the PCM is the 
low momentum transfer cut-off $p_T^{min}$. Under certain assumptions 
this parameter can be determined from experimental data 
for elementary hadron-hadron collisions \cite{sjostrand,naga,eskola1}.
In the environment of a heavy-ion collision, colour
screening will destroy the association of partons to particular hadrons,
since for a sufficiently high density of colour charges, the colour
screening radius becomes much smaller than the typical hadronic scale.
It is therefore by no means clear whether the $p_T^{min}$ values 
extracted from hadron-hadron collisions are applicable to heavy-ion
collisions. Calculating the colour screening mass $\mu_D$ for
a set of systems where interactions are governed by  specific values 
of $p_T^{min}$ may allow us to determine a lower boundary for the allowed
range of $p_T^{min}$ values, since only values of $p_T^{min} \ge \mu_D$
are physical.  

Following ref. \cite{BMW}, we use perturbative QCD
to obtain the
time evolution of the screening mass $\mu_D(\tau)$.
The parton cascade model provides the phase space distribution of the 
partons. The general form for the colour screening 
mass in the one loop approximation is  \cite{BMW,sspc,klimov}
\begin{widetext}
\begin{equation}
\label{mud_form}
\mu_D^2=-\frac{3\alpha_s}{\pi^2}\,
      \lim_{|\vec{q}|\rightarrow 0}
      \int\, d^3p\,
       \frac{|\vec{p}|}{\vec{q}\cdot\vec{p}}
     \,  \vec{q}\cdot\nabla_{\vec{p}}
   \left[ F_{g}(\vec{p})+\frac{1}{6}\sum_q
    \left\{ F_q(\vec{p})+F_{\overline{q}}(\vec{p})\right\} \right],
\end{equation}
\end{widetext}
where $\alpha_s$ is the strong coupling constant, the $F_i$ specify
the phase space density of gluons, quarks, and anti-quarks and $q$
runs over the flavour of quarks. It is easy
to verify that in the case of an ideal gas of massless partons, where the
$F_i$ reduce to Bose-Einstein or Fermi-Dirac distributions (with
vanishing baryochemical potential $\mu_B$), Eq.~\ref{mud_form} reduces 
to the standard result for the thermal Debye mass \cite{klimov,weldon}.
The partonic
distribution will be initially anisotropic with respect to the beam axis
and thus the screening mass of a gluon in
the matter may depend on the direction of propagation.
We have found that $\mu_D^\perp$ differs from $\mu_D^{||}$ by 10\% at most
and we have therefore assumed $\mu_D = \mu_D^\perp$ in the following 
discussion.
We also note that the assumptions underlying this method
are not strictly applicable to  very early times, $\tau < \Delta z$, where
$\Delta z$ is the Lorentz contracted width of the nuclei.

\begin{figure}[t]
\includegraphics[width=3.5in]{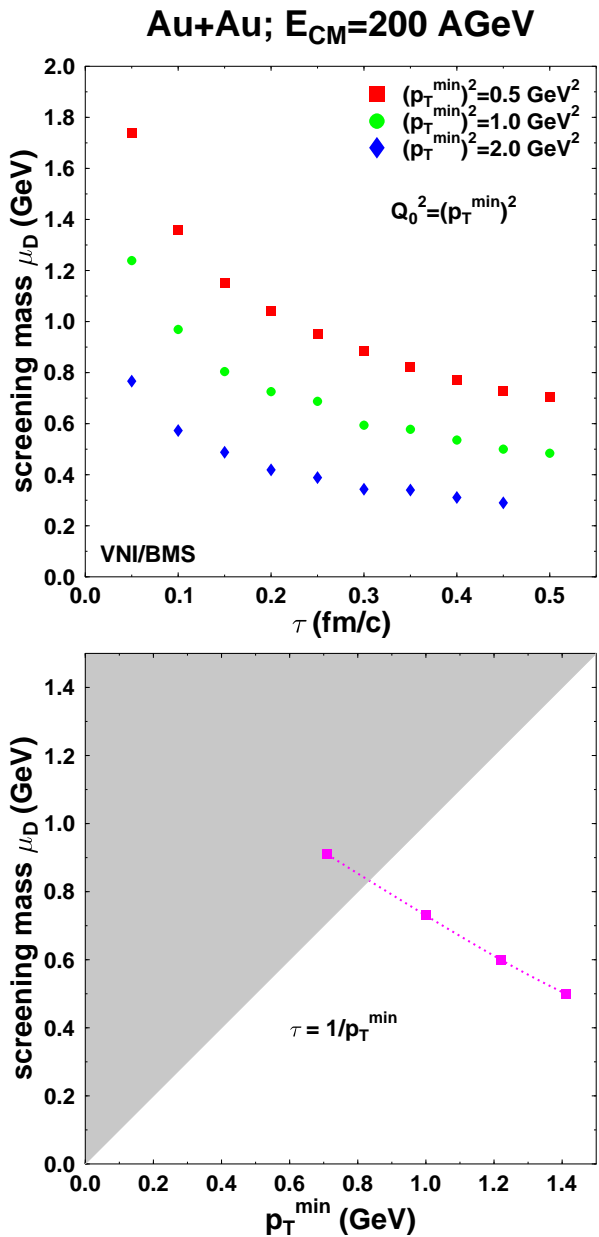}
\caption{\label{mud1} Top: time-evolution of the screening mass $\mu_D$ 
for several values of $p_T^{min}$ in Au+Au collisions at RHIC. 
Bottom: $\mu_D$ evaluated at 
$\tau=1/p_T^{min}$ as a function of $p_T^{min}$. The gray shaded area symbolizes
the unphysical region in which $p_T^{min} \le \mu_D$ and therefore serves
as a boundary for the allowed range of $p_T^{min}$ in Au+Au collisions at 
RHIC. }
\end{figure}

\begin{figure}[t]
\includegraphics[width=3.5in]{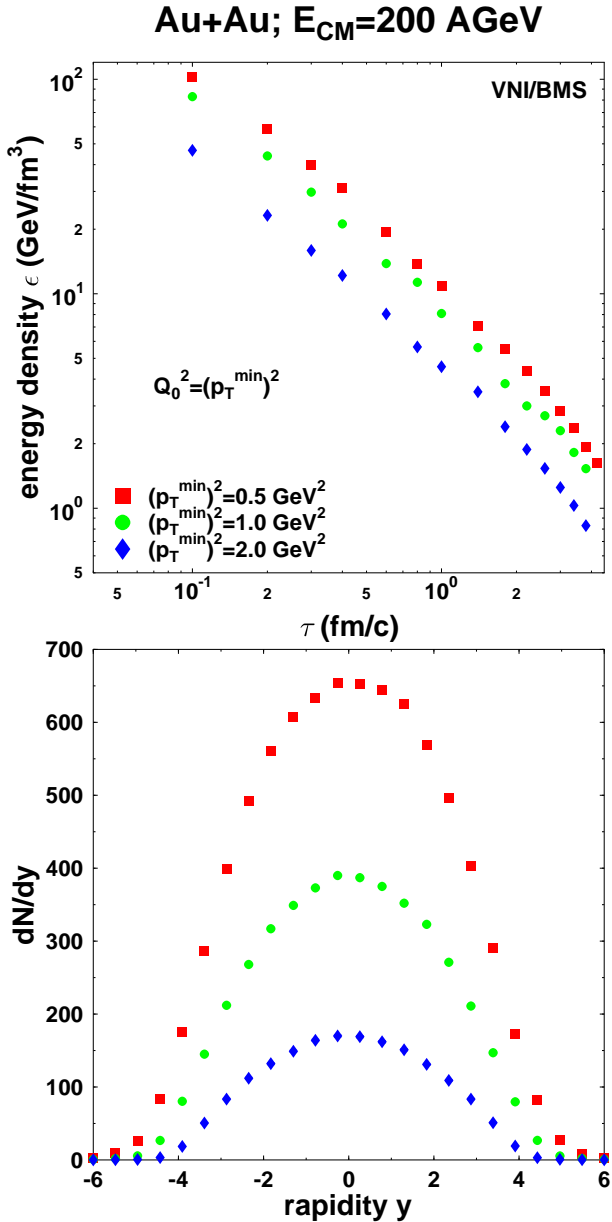}
\caption{\label{eps_t} Top: time-evolution of the energy-density
for different values of $p_T^{min}$. Bottom: rapidity distribution
of produced partons generated by hard and semi-hard parton-parton
interactions at RHIC.} 
\end{figure}

\begin{figure}[th]
\includegraphics[width=3.5in]{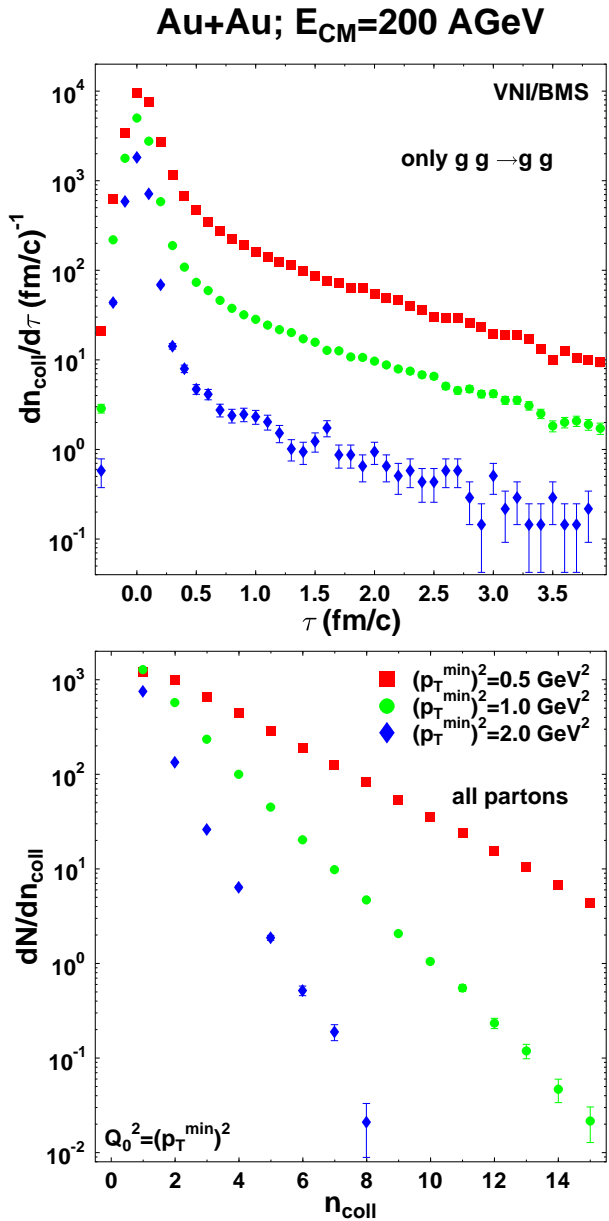}
\caption{\label{coll} Top: collision rates for gluon-gluon scattering
as a function of time for different values of $p_T^{min}$. Bottom: 
parton collision number distribution for different $p_T^{min}$: the
probability for a parton to suffer multiple collisions increases 
dramatically with decreasing $p_T^{min}$.}
\end{figure}

We begin by calculating the screening mass for partonic matter in 
central Au+Au collisions at RHIC ($\sqrt{s}=200$~GeV per nucleon pair). The
upper frame of figure~\ref{mud1} shows the time evolution of the 
(transverse) screening mass, calculated according to eq.~\ref{mud_form}
for several different values of $p_T^{min}$.
The time $t=0$ marks the maximum
overlap of the two colliding nuclei.  
The time-evolution of $\mu_D$ clearly reflects the dynamics of
the collision: at early times the density of the system 
is large, leading to a large value  the screening mass $\mu_D$. 
For later times, density and collision rate (see also figures~\ref{eps_t}
and~\ref{coll}) decrease, reducing the value of $\mu_D$ by more than a 
factor of two. 
The strong time-dependence of the screening mass during
the early pre-equilibrium phase will have an influence on the 
equilibration process that is beyond the scope of our present work. 
A decreasing screening mass implies a 
rising cross section and thus enhanced multiple rescattering.
This aspect is missing in many implementations of the PCM, which assume
a time-independent screening mass for the entire duration of the collision
\cite{zhang,molnar,ampt,gromit}. 

A self-consistent calculation would utilize the time-dependent
Debye-mass as the regulator of the interaction among partons instead of 
introducing a fixed cut-off $p_T^{min}$ as used in our present work. 
This would be achieved by calculating the parton-parton cross sections
in the framework of the hard thermal loop (HTL) approximation \cite{BP90},
which accounts for the full frequency and momentum dependence of the
dynamic screening in the high-density, small gradient limit of QCD.
The implementation of such a self-consistent dynamical screeing 
mechanism, as formulated recently by Arnold et al. \cite{AMY02} 
in a PCM remains a major challenge for future work.
It should also be noted that, strictly speaking, newly produced 
partons start providing screening only after  $\Delta \tau \sim 1/p_T$
and therefore  the concept of a screening mass may not be well justified 
for times $\tau \le 1/p_T^{min}$ \cite{sspc}. 

In order to determine a lower boundary for the allowed range of  $p_T^{min}$,
we calculate $\mu_D$ at $\tau = 1/p_T^{min}$ and plot it as a function
of $p_T^{min}$ in the bottom frame of figure~\ref{mud1}.
The gray shaded area symbolizes
the region in which $p_T^{min} \le \mu_D$ and where the procedure to simply
cut off the interaction is no longer valid.
It therefore serves
as a boundary of the allowed range of $p_T^{min}$ for our calculation.
We find that the smallest allowed value for $p_T^{min}$ is $\approx 0.8$~GeV.
The inclusion of higher order radiative corrections and parton fusion 
processes into our calculation will probably 
alter this value -- we shall study these
effects in a forthcoming publication \cite{bms_big1}.

After having determined the meaningful range of $p_T^{min}$, 
our model is applied to the calculation of the time-evolution
of the energy-density $\epsilon$ generated by hard and semi-hard parton-parton
interactions at RHIC: the upper frame of figure~\ref{eps_t} shows
this time-evolution of $\epsilon$ for different values of $p_T^{min}$, 
to give an estimate on the sensitivity of $\epsilon$ with respect to
$p_T^{min}$. Here, $\epsilon$ is calculated via
\begin{equation}
\epsilon (r_T) \, = \, \frac{1}{2 \pi r_T \tau} 
\left(\frac{{\rm d}^2E_T(r_T)}{{\rm d}y\,{\rm d}r_T}\right)_{y=y_{CM}}
\end{equation}
and
\begin{equation}
\langle \epsilon \rangle \,=\, \frac{1}{\pi R_T^2}\,
\int_{0}^{R_{T}} \epsilon (r_T) \, 2 \pi r_T \, {\rm d} r_T
\end{equation}
and choosing  $R_{T}=2$~fm.
The maximum energy density obtained in Au+Au collisions in the PCM approach
is found to be on the order of 100~GeV/fm$^3$.
A detailed analysis shows that for times $\tau \le 1.0$~fm/c $\epsilon(\tau)$
scales with $1/\tau$ whereas for later times the scaling function is
$1/\tau^{4/3}$, most likely signaling 
a transition from longitudinal streaming
to a three dimensional expansion. 
One might be tempted to attribute this scaling to the onset of hydrodynamic
expansion. However, the small collision rates at later times (see below)
preclude this interpretation.

The bottom frame of figure~\ref{eps_t} shows the rapidity distribution
$dN/dy$ of produced partons for different values of $p_T^{min}$.
While the distributions cannot be directly compared to experimental data,
they provide a lower limit on the entropy produced via hard and semi-hard
interactions. Since at hadronization entropy can only be produced or 
remain constant, the rapidity density of the produced partons provides
us with a lower bound for the rapidity density of hadrons compatible  
with the parameters of our calculation. For $p_T^{min}=0.7$~GeV the
rapidity density of produced partons at mid-rapidity is approximately 650.
Assuming parton-hadron duality and entropy conservation during
hadronization this number would translate to approximately 430 charged
hadrons. In order to be compatible with the data \cite{rhic_data},
it is important for
this number to remain well below the measured value, to allow for
additional contributions due to intial- and final state radiation as well
as soft particle production.

\begin{figure}[t]
\includegraphics[width=3.5in]{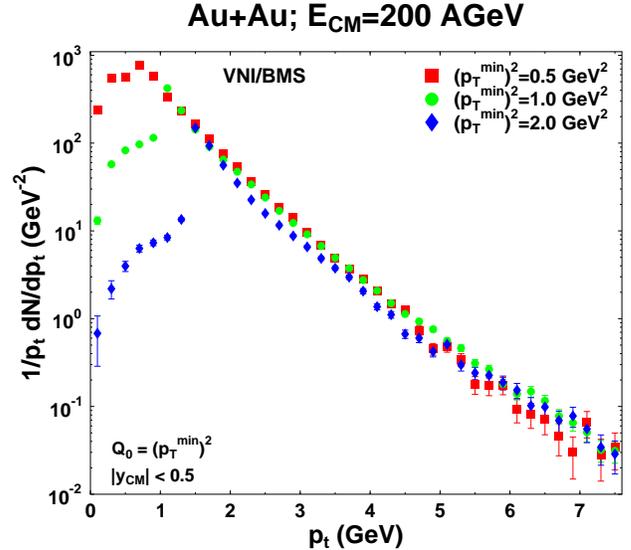}
\caption{\label{ptspec} Transverse momentum spectrum of produced partons
in central Au+Au collisions at RHIC for different values of $p_T^{min}$. 
Note that the contributions to the spectrum below $p_T^{min}$ stem from
parton rescattering.}
\end{figure}

While the absolute values for the energy and rapidity density 
 show a significant dependence on $p_T^{min}$, the overall shape
of the transverse momentum distribution of produced partons changes
only in a subtle manner, as can be seen in figure~\ref{ptspec}. The main
impact the choice of $p_T^{min}$ has, is in the low-$p_T$ starting point
of the respective distribution -- contributions below $p_T^{min}$ stemming
from parton rescattering. Naively one would think that the transverse
momentum distribution above the chosen cut-off should not be affected by
the choice of the cut-off -- however, choosing a low cut-off value
increases the amount of parton rescattering which 
visibly modifies the shape of the $p_T$ spectrum.

In order to quantify the amount of parton scattering and rescattering
we evaluate the collision rates as a function of time and $p_T^{min}$.
The upper frame of figure~\ref{coll} shows the collision rates 
for gluon-gluon scattering (glue-glue collisions account for roughly
half the total number of hard scatterings). The time-evolution of the
collision rate is clearly peaked at $\tau=0$~fm/c when the maximum overlap
between the two gold nuclei occurs. For $-0.5 \le \tau \le 0.5$~fm/c
the collision rates show a rough symmetry around $\tau=0$~fm/c for 
large values of $p_T^{min}$, indicating
the dominance of primary-primary parton scattering contributions to the
collision rate during early times if the cross section is small. This
symmetry gets distorted for decreasing $p_T^{min}$, when secondary
parton rescattering increases.  
For later times the steep decrease in the collision rates
flattens out due to parton rescattering, exhibiting a far larger
sensitivity on $p_T^{min}$ (and thus the interaction cross section) than
at early times. The probability for a parton scattering multiple times
can be directly evaluated by plotting the parton collision number
distribution, which can be seen in the lower frame of figure~\ref{coll}.
While for all cases it is most likely for a parton to only undergo one
hard or semi-hard collision, the
probability for a parton to suffer multiple collisions increases 
dramatically with decreasing $p_T^{min}$.

In a future analysis, we shall attempt to isolate those regions in
phase-space in which rescattered partons dominate
and identify experimentally
accessible observables which are most sensitive to multiple binary
perturbative rescattering of partons.

In summary, we have calculated the time-evolution of the 
Debye-screening mass $\mu_D$
for Au+Au collisions at RHIC in the framework of a lowest-order 
Parton Cascade Model.
We have used
the screening mass to determine a lower boundary for the allowed
range of $p_T^{min}$ values. For that range
we determined the energy density reached through hard and 
semi-hard processes and found that its maximum remains below 
approx. 70 GeV/fm$^3$.
We established  a lower bound for the rapidity
density of charged hadrons which is  compatible with the available data. 
The possibility of perturbative rescattering among partons has been 
analyzed and is found to significantly increase with decreasing $p_T^{min}$.

\begin{acknowledgments} 
This work was supported by  RIKEN, Brookhaven
National Laboratory and DOE grants DE-FG02-96ER40945 as well as
DE-AC02-98CH10886. One of us (DKS) is especially grateful for the warm
and generous hospitality at Duke University, where a large part of this
work was done.
\end{acknowledgments}

\end{document}